\documentclass[traditabstract]{aa} 
                                   
\bibpunct{(}{)}{;}{a}{}{,}

\newcommand{\ergps}{erg\thinspace s$^{-1}$}

\newcommand{\kmps}{km\thinspace s$^{-1}$}
\newcommand{\psqcm}{cm$^{-2}$}
\newcommand{\nH}{$N_{\rm H}$}

\newcommand{\lya}{Ly$\alpha $}

\usepackage[varg]{txfonts}
\usepackage{graphicx}
\usepackage{subfig}
\usepackage{hyperref}
\begin{document}
\title{Subaru High-z Exploration of Low-Luminosity Quasars (SHELLQs)}

\subtitle{XXII. Chandra observations of narrow-line quasar candidates at $z\geq 6$}

\author{K. Iwasawa\inst{1,2}
\and
R. Gilli\inst{3}
\and
F. Vito\inst{3}
\and
Y. Matsuoka\inst{4}
\and
M. Onoue\inst{5,6}
\and
M. A. Strauss\inst{7}
\and
N. Kashikawa\inst{8}
\and
Y. Toba\inst{9,10}
\and
K.~Shimasaku\inst{8}
\and
K. Inayoshi\inst{11}
\and
T. Nagao\inst{4}
\and
N. Kawanaka\inst{12,13}
\and
J.~D.~Silverman\inst{6}
\and
T. Izumi\inst{13}
\and
K.~Kohno\inst{14,15}
\and
Y.~Ueda\inst{16}
}

\institute{Institut de Ci\`encies del Cosmos (ICCUB), Universitat de Barcelona (IEEC-UB), Mart\'i i Franqu\`es, 1, 08028 Barcelona, Spain
         \and
ICREA, Pg. Llu\'is Companys 23, 08010 Barcelona, Spain
\and
INAF – Osservatorio di Astrofisica e Scienza dello Spazio di
Bologna, Via Gobetti 93/3, I-40129 Bologna, Italy
\and
Research Center for Space and Cosmic Evolution, Ehime University, 2-5 Bunkyo-cho, Matsuyama, Ehime 790-8577, Japan
\and
Waseda Institute for Advanced Study (WIAS), Waseda University, 1-21-1, Nishi-Waseda, Shinjuku, Tokyo 169-0051, Japan
\and
Kavli Institute for the Physics and Mathematics of the Universe (Kavli IPMU, WPI), UTIAS, Tokyo Institutes for Advanced Study, University of Tokyo, Chiba, 277-8583, Japan
\and
Department of Astrophysical Sciences, Princeton University, 4 Ivy Lane, Princeton, NJ 08544, USA
\and
Department of Astronomy, School of Science, The University of Tokyo, 7-3-1 Hongo, Bunkyo, Tokyo 113-0033, Japan
\and
Department of Physical Sciences, Ritsumeikan University,Kusatsu Shiga 525-8577, Japan
\and
Academia Sinica Institute of Astronomy and Astrophysics, 11F Astronomy-Mathematics Building, AS/NTU, No.1, Section 4, Roosevelt Road, Taipei 10617, Taiwan
\and
Kavli Institute for Astronomy and Astrophysics, Peking University, Beijing 100871, China
\and
Department of Physics, Graduate School of Science Tokyo Metropolitan 
University, Hachioji-shi, Tokyo 192-0397
\and
National Astronomical Observatory of Japan, Osawa, Mitaka, Tokyo 181-8588, Japan
\and
Institute of Astronomy, Graduate School of Science, The University of Tokyo, 2-21-1 Osawa, Mitaka, Tokyo 181-0015, Japan
\and
Research Center for the Early Universe, Graduate School of Science, The University of Tokyo, Hongo, Tokyo 113-0033, Japan
\and
Department of Astronomy, Kyoto University, Sakyo-ku, Kyoto, Japan
}
 
\abstract{
  We report on Chandra X-ray observations of four narrow-line quasar candidates at $z\sim 6$, selected from the SHELLQs project, based on the Subaru Hyper Suprime-Cam survey. These objects are characterised by narrow (FWHM$\leq 310$ km\,s$^{-1}$), luminous ($>10^{44}$ \ergps) Ly$\alpha $ and faint UV continuum ($M_{1450} = -22$ - $-21$), prompting us to examine whether they are obscured luminous AGN at the epoch of reionization. However, none of these objects were detected by Chandra, giving an upper limit to their rest-frame 2-10 keV luminosity ($L_{\rm X}$) of $2\times 10^{44}$ \ergps\ ($2\sigma$), assuming a spectral slope $\Gamma=2$.
Subsequent rest-frame optical spectroscopy of these objects by the JWST-NIRSpec, presented in a companion paper, show weak broad Balmer emission at the base of narrow cores. With the scaling relation for low-redshift AGN, the observed strong [O{\sc iii}]$\lambda 5007$ flux of these sources would predict $L_{\rm X}$ to be around $10^{45}$ \ergps, which is well above the Chandra upper limits. These optical spectra and X-ray quietness are reminiscent of JWST-selected broad-line AGN. We attribute the weak broad Balmer emission to the broad-line regions hidden partially by optically-thick obscuring matter which also hides the optical and X-ray continuum emission from the accretion disc. Compton-thick obscuration, which would strongly suppress X-ray emission, could be due to a dense inter-stellar medium that is often present in galaxies at high redshifts. Alternatively, the same effect could be obtained from an inflated disc at the innermost radii in a supercritical accretion flow, when the disc is viewed at inclined angles. 
}

\titlerunning{X-ray emission from SHELLQs narrow-line objects}
\authorrunning{K. Iwasawa et al.}
\maketitle
%

\section{Introduction}

Both X-ray surveys and theoretical studies indicate that obscuration of active galactic nuclei (AGN) increases towards higher redshift, and the obscured AGN fraction at $z>4$ could be as high as 80-90\% \citep{gilli2022,ni2020,davies2019,trebitsch2019,vito2018,liu2017}.
X-ray observations prove to be an useful tool in searching for obscured AGN, owing to the penetrating nature of X-rays \citep[e.g.][]{hickox2018}. The search for obscured quasars at high redshift is helped by the negative k-correction of an absorbed X-ray spectrum, unless the obscuration is heavily Compton-thick. In the absence of broad emission lines, optical-UV AGN signatures indicating a powerful ionising source, together with X-ray detection, would give a convincing case for obscured AGN. Despite the expected abundance of obscured AGN at high redshift, few examples showing both AGN narrow emission-line features and powerful X-ray emission have been found.

\begin{figure}
  \centerline{\includegraphics[width=0.5\textwidth,angle=0]{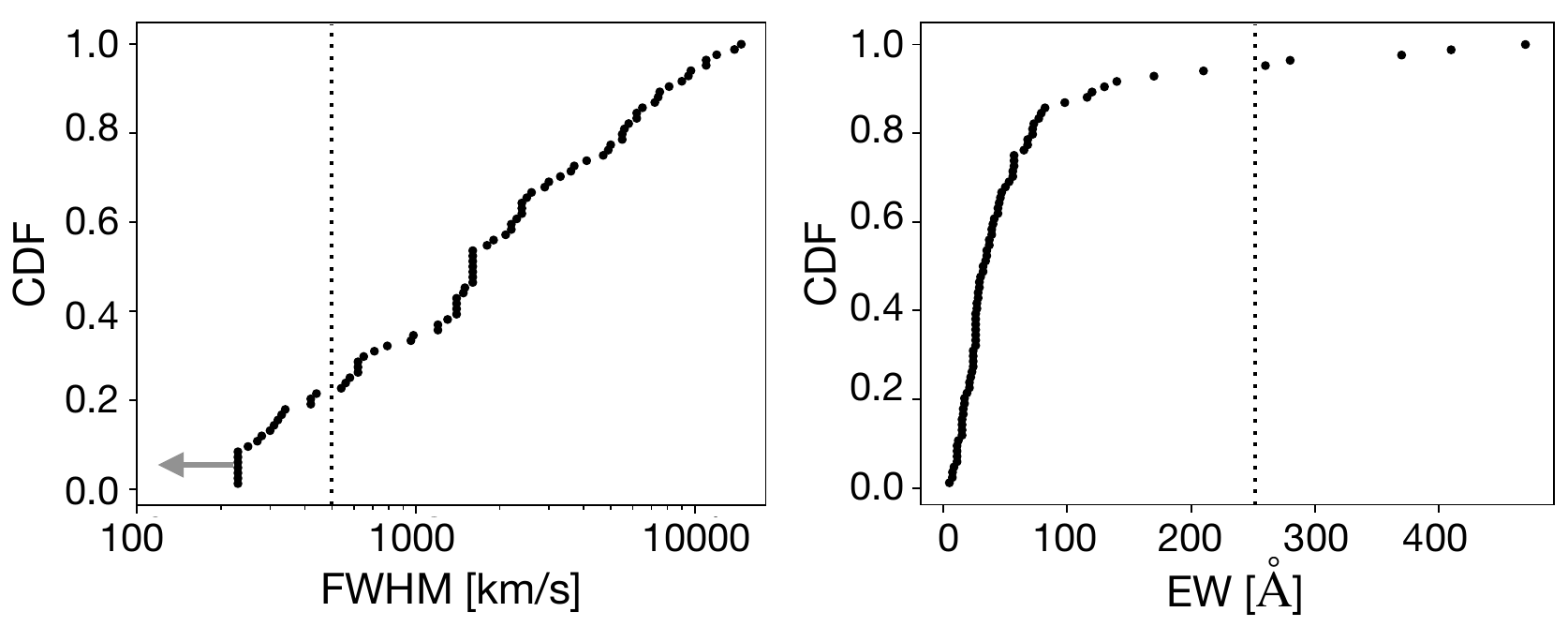}}
  \caption{Line widths (FWHM) and rest-frame equivalent width (EWs) of Ly$\alpha $ of SHELLQs quasars. Left panel: CDF of FWHM(Ly$\alpha $) in units of km\, s$^{-1}$. Seven objects show unresolved lines that have an upper limit on FWHM of 230 km\, s$^{-1}$. Dotted line indicates the dividing line at 500 km\,s$^{-1}$ below which narrow-line objects are defined. Right panel: CDF of EW(Ly$\alpha $). Dotted line indicates 250 \AA, above which five outliers in Fig. \ref{fig:LM_diagram}, including the four Chandra targets are found.}
\label{fig:CDFs}
\end{figure}

The ``Subaru High-z Exploration of Low-Luminosity Quasars'' (SHELLQs) project \citep{matsuoka2016} searches for quasars in the redshift range of $z >5.6$, based on the Hyper Suprime-Cam \citep[HSC,][]{miyazaki2018} Subaru Strategic Program (SSP) wide-field imaging survey \citep{aihara2018}. The majority of these quasars lie in the intermediate luminosity range between those discovered by earlier works based on shallower surveys \citep[e.g.][]{fan2006,willott2005,jiang2015} and faint broad-line AGN including Little Red Dots (LRDs), recently found by James Webb Space Telescope (JWST) spectroscopy \citep[e.g.][]{harikane2023,ubler2023,greene2024,matthee2024,maiolino2024,juodzbalis2024,wang2024,schindler2024,furtak2024,kocevski2024}. The SHELLQs quasars in this luminosity range are quite rare, requiring a sensitive survey over the large area $\sim 1000$ deg$^{2}$ of the Subaru HSC SSP to be discovered. The SHELLQs quasars include 18 faint objects with luminous narrow Ly$\alpha $ \citep{matsuoka2018apjs,matsuoka2018pasj,matsuoka2019}. They are characterised by their Ly$\alpha $ line widths and luminosities: FWHM\,$< 500$ \kmps\ [see the cumulative density function (CDF) of FWHM(Ly$\alpha $) in Fig. \ref{fig:CDFs}] and $L$(Ly$\alpha )>1\times 10^{43}$ \ergps\ \citep{matsuoka2018pasj}. It has been known that Ly$\alpha $ emitters (LAEs) exceeding this luminosity are dominated by AGN \citep{konno2016,sobral2018}. Similar narrow-line quasar candidates have been discovered at lower redshifts \citep[e.g.][]{alexandroff2013}, but few were known at $z\sim 6$ \citep{willott2009,kashikawa2015}. Some of the SHELLQs narrow-line objects have $L$(Ly$\alpha $) exceeding $10^{44}$ \ergps, more luminous than any known star-forming LAEs, whereas they show faint UV continuum emission, consistent with galaxy light with $M_{1450}\geq -22$. These objects can be considered as good obscured AGN candidates.
Keck-MOSFIRE near-infrared spectroscopy of one of the narrow-line objects, J142331.71--001809.1, revealed narrow C{\sc iv} doublet $\lambda\lambda 1548,1550$ with a rest-frame equivalent width, EW$\simeq 37$\AA\ \citep{onoue2021}. Since the maximum EW of C{\sc iv} expected in star forming galaxies is 10\AA\ \citep{nakajima2018}, this strongly suggests the presence of hard ionising photons from an AGN.

Here we describe X-ray observations of four of these narrow-line objects with the Chandra X-ray Observatory (Chandra hereafter), which resulted in no detection. Combined with results of recently acquired JWST rest-frame optical spectra of these objects which show broad Balmer lines \citep{matsuoka2025}, we conclude that these objects may be higher-luminosity versions of the JWST-selected AGN, as they share similar properties. This paper is structure as follows: Sect. 2 describes the target selection, followed by the Chandra observations (Sect. 3) and the results (Sect. 4). Sect. 5 discusses implications of the results, including the JWST spectroscopy of the targets. We adopt the cosmological parameters $H_0=70$\,km\,s$^{-1}$\,Mpc$^{-1}$, $\Omega_{\rm M}=0.3$, and $\Omega_{\Lambda}=0.7$. 

\section {Target selection}

\subsection{Ly$\alpha $ properties}
  
\begin{figure}
\centerline{\includegraphics[width=0.4\textwidth,angle=0]{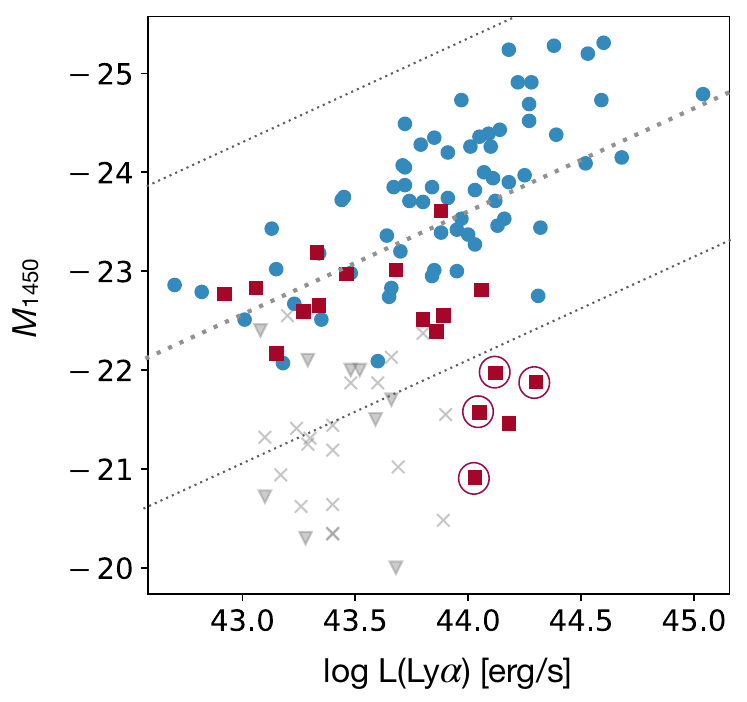}}
\caption{UV absolute magnitude $M_{1450}$ as a function of Ly$\alpha $ luminosity of SHELLQs quasars. Broad-line (FWHM(Ly$\alpha $)$\geq 500$ km\,s$^{-1}$) and narrow-line (FWHM(Ly$\alpha$)$<500$ km\,s$^{-1}$) objects are plotted with blue circles and red squares, respectively. The four Chandra targets are circled in red. Dashed line indicates the best linear fit to both broad- and narrow-line objects. The two dotted lines show the 95\% compatible intervals. Data for LAEs at $z\sim 6$ detected in the Subaru HSC survey from \citet{shibuya2018} are also plotted in grey for comparison. Crosses are LAEs for which UV continuum emission was detected while triangles are those for which only upper limits of UV emission were obtained. }
\label{fig:LM_diagram}
\end{figure}

We selected Chandra targets from among the 18 SHELLQs narrow-line quasar candidates discovered as of \citet{matsuoka2019}. Fig. \ref{fig:LM_diagram} show the $M_{1450}$\,-\,$L$(Ly$\alpha $) diagram of SHELLQs quasars, divided into broad-line (FWHM\,$\geq 500$ \kmps) and narrow-line objects (Fig. \ref{fig:CDFs}). A loose correlation between the line and continuum luminosities can be seen. There are five objects that lie more than $2\sigma $ from the linear fit to $M_{1450}$ against log $L({\rm Ly}\alpha )$ for the full SHELLQs sample\footnote{If these five points are excluded, we find a steeper correlation between $M_{1450}$ and log $L$(Ly$\alpha $), and a scatter reduced by 30\%}.
The five outliers happen to have the largest Ly$\alpha $ luminosities among the narrow-line objects. The high EW of Ly$\alpha $ is consistent  with a suppressed UV continuum in an obscured AGN. As shown in Fig. \ref{fig:LM_diagram}, they might be objects at the extreme end in luminosity of the population of luminous LAEs at $z\sim 6$ detected in the Subaru HSC survey \citep{shibuya2018}, which could also contain AGN. We thus selected these five objects as possible obscured AGN candidates.

Due to the limited observing time awarded for this program, one of the candidates, J114658.96--000537.6 ($z=6.30$)
was left out and the other four objects were observed with Chandra. Table \ref{tab:targets} gives rest-frame UV properties of the four targets. We refer these objects as G01, G02, G03 and G04 hereafter, as given in the companion paper on JWST results by \citet{matsuoka2025}. They have \lya\ emission with luminosities exceeding $10^{44}$ \ergps\ and EW ranging between 280\AA\ and 470\AA\ in the rest frame (Table \ref{tab:targets}, Fig. \ref{fig:CDFs}).

\begin{figure}
\centerline{\includegraphics[width=0.4\textwidth,angle=0]{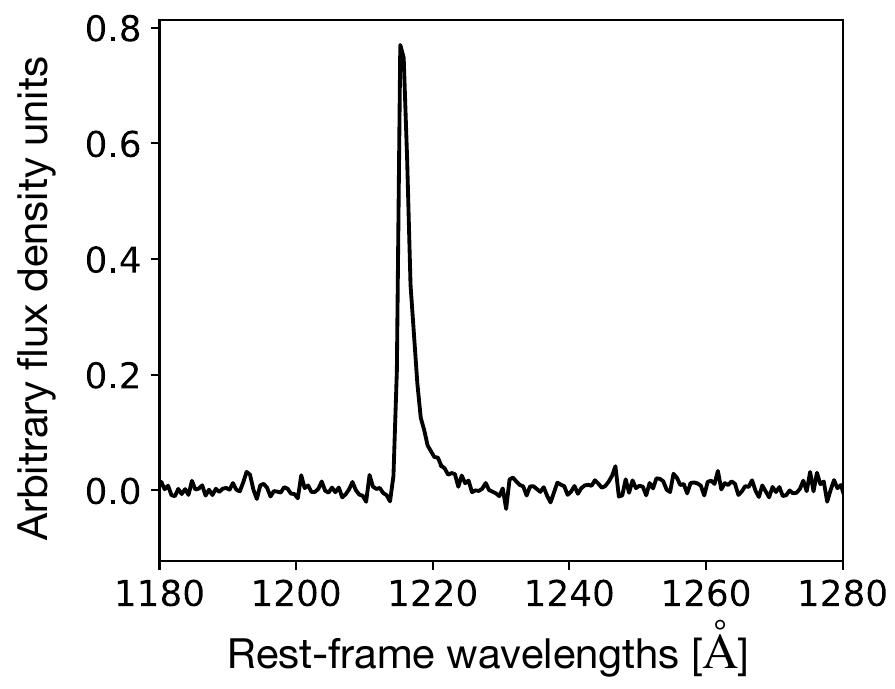}}
\caption{Mean Ly$\alpha $ profile of the four Chandra targets.}
\label{fig:lyaprofile}
\end{figure}

The mean line profile of Ly$\alpha $ of the four objects, obtained from the Subaru-FOCAS \citep{matsuoka2019}, which was constructed after normalising the spectra by the line peak, is shown in Fig. \ref{fig:lyaprofile}. The line profile of this composite shows a weak red-wing (the line profile bluewards is absorbed by the Ly$\alpha $ forest). This wing component has a FWHM of $\sim 2000$ km\,s$^{-1}$ (detailed fits of the line profiles of individual objects are presented in \citet{matsuoka2025}). Although this broad component could be attributed to the BLR of AGN, other causes such as an outflow and resonant scattering \citep[e.g.][]{osterbrock1962} might be in play. 
We note that there is no clear sign of N{\sc v}$\lambda 1240\AA $ with a P Cygni profile, seen in the composite spectrum of all SHELLQs narrow-line objects presented in \citet{matsuoka2019}. 

The observed UV continuum luminosities of these objects are all fainter than $M_{1450} = -22$, and thus can be attributed to galaxy emission alone, as the number density of galaxies is about three orders of magnitude above that of unobscured quasars at $z\sim 6$, suggested by their respective luminosity functions \citep{matsuoka2018lf}.

\begin{table*}
  \caption{Chandra targets}
  \label{tab:targets}
  \centering
  \begin{tabular}{ccccccc}
    \hline\hline
    Name & Object & $z$ & FWHM & EW & log\,$L$(Ly$\alpha $) & $M_{1450}$ \\
    & & & \kmps & \AA & [\ergps] & mag \\
    \hline
    G01 & J142331.71--001809.1 & 6.127 & $<230$ & 370 & 44.30 & $-21.88$ \\
    G02 & J084456.62+022640.5 & 6.397 & $<230$ & 280 & 44.05 & $-21.57$ \\
    G03 & J093543.32--011033.3 & 6.075 & $<230$ & 410 & 44.12 & $-21.97$ \\
    G04 & J125437.08--001410.7 & 6.130 & $310$ & 470 & 44.03 & $-20.91$ \\
    \hline
    \end{tabular}
  \tablefoot{Short target names, G01..G04, used in this work are identical to those used in \citet{matsuoka2025}. Object names follow the Subaru HSC survey naming convention and convey right ascension and declination in J2000 coordinates for each source. The redshifts shown here are updated values, obtained from the JWST-NIRSpec spectra \citep{matsuoka2025}. FWHM is the full width at half maximum of Ly$\alpha $ as measured, without correction for absorption of the intergalactic medium. EW is the rest-frame equivalent width of Ly$\alpha $. $M_{1450}$ is absolute magnitude at 1450\,\AA.}
\end{table*}

\subsection{X-ray detection experiment}

The observation of the \lya\ emission line was the only clue we had before the X-ray observations for how powerful possible AGN in these sources might be. Its luminosity can be a rough guide of AGN luminosity: although statistically there is a positive correlation between the luminosities of \lya\ and X-ray in AGN \citep{calhau2020}, the scatter of the correlation and sparse sampling of the large Ly$\alpha $ luminosity range mean that $L({\rm Ly}\alpha )$ is not a reliable $L_{\rm X}$ predictor for an individual object. We therefore took an alternative route and predicted their $L_{\rm X}$ based on their surface density, as described below, and designed the Chandra observations as a detection experiment in which only a few source counts are expected.

Five obscured AGN candidates which are outliers of the $M_{1450}$-$L({\rm Ly}\alpha )$ correlation (Sect. 2.1) were found in the survey area of 900\,deg$^2$. Suppose obscured AGN outnumber unobscured AGN by a factor of $\beta $, and the obscuration-corrected luminosity function of obscured AGN has the same shape as that of the known quasars at $z\sim 6$ of \citet{matsuoka2018lf}. We assume $\beta $ to be in the range of 1 to 10, which is equivalent to an AGN obscured fraction $f_{\rm obsc} = 0.5$-0.91. The lower bound corresponds to the limit of obscuration caused only by circumnuclear hot dust found in quasars at $z\sim 2$ \citep{lusso2013}. The higher bound matchs the value $f_{\rm obsc}>0.82$ derived from the proximity zone analysis for $z>7$ quasars by \citet{davies2019}. Given this hypothetical luminosity function, you can obtain a luminosity that yields the observed surface density. For the two extreme values of $\beta $, the UV absolute magnitudes to yield five objects per 900\,deg$^2$ are $M_{1450}\simeq -25.7$ ($\beta = 1$) and $M_{1450}\simeq -27.7$ ($\beta = 10$). Using the known luminosity-dependent UV to X-ray relation \citep{vito2019}, the corresponding rest-frame 2-10 keV luminosities, $L_{\rm X}$, are $4\times 10^{44}$ \ergps\ ($\beta = 1$) and $1.3\times 10^{45}$ \ergps ($\beta = 10$). If the targets emit at luminosity in the above range and have a power-law spectrum of photon index $\Gamma = 2$, source counts expected in the 1-5 keV band for a 60 ks Chandra ACIS-S observation range from five to 15 counts for $z=6.1$, even if the equivalent hydrogen absorbing column density of obscuration is as large as \nH\,$\sim 10^{23}$ \psqcm.

\section{Observations}

Chandra observations of the four SHELLQs narrow-line objects were carried out during the Cycle 23. Each target was observed with three or four exposures to achieve an integrated exposure time of approximately 60 ks, as shown in Table \ref{tab:obslog}. All the targets were placed on the ACIS-S3 chip, operating in the VFAINT mode. The Galactic absorption for each target ranges from \nH\,$=1.4\times 10^{20}$ \psqcm\ to \nH\,$=3.9\times 10^{20}$ \psqcm, according to the HI4PI map (HI4PI collaboration, 2016). The Chandra ACIS detector of the current effective area is not sensitive to this amount of small Galactic absorption. 
Data reduction was performed in the standard manner using CIAO\,4.15 \citep{fruscione2006}.

\begin{table}
\caption{Chandra observation log}
\label{tab:obslog}
\centering
\begin{tabular}{lccc}
  \hline\hline
Name & ObsID & Date & Exposure \\
\hline
  G01 & 25370 & 2021-12-14 & 19.82 \\
  & 25905 & 2023-05-07 & 27.11 \\
  & 26237 & 2021-12-15 & 9.82 \\
  G02 & 25371 & 2023-01-26 & 14.74 \\
  & 25894 & 2022-02-03 & 13.90 \\
  & 26300 & 2022-02-04 & 14.76 \\
  & 27675 & 2023-01-27 & 13.20 \\
  G03 & 25373 & 2023-02-06 & 15.86 \\
  & 25740 & 2022-10-10 & 11.95 \\
  & 27478 & 2022-10-12 & 15.68 \\
  & 27699 & 2023-02-06 & 13.71 \\
  G04 & 25372 & 2022-04-06 & 13.77 \\
  & 25862 & 2023-04-10 & 14.88 \\
  & 26379 & 2022-04-06 & 16.85 \\
  & 27797 & 2023-04-10 & 14.88 \\
  \hline
\end{tabular}
\tablefoot{Exposure time is in units of $10^3$ seconds. All the observations were carried out with the ACIS-S3 operating in VFAINT mode.}
\end{table}

\section{Results}

\begin{figure}
\centerline{\includegraphics[width=0.43\textwidth,angle=0]{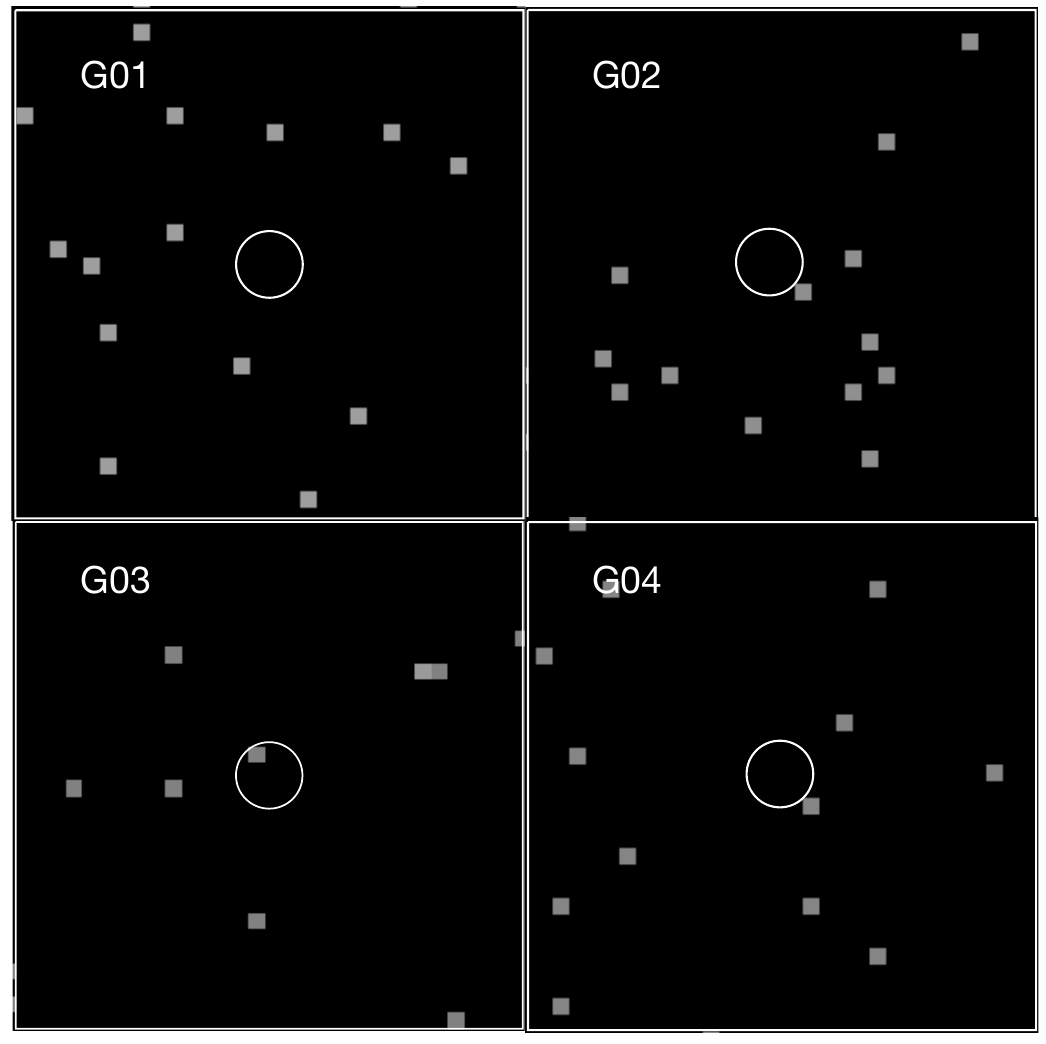}}
\caption{Chandra 1-5 keV images of the four narrow-line SHELLQs quasars. Each panel is a 15\arcsec $\times $15\arcsec\ field of view centred on the target. The image orientation is North up, East to the left. The circle indicates the optical position with a 1 arcsec radius. }
\label{fig:e15imgs}
\end{figure}

X-ray images of each target were constructed by stacking the repeat exposures. The total exposure times are and 56.7 ks for G01, 56.5 ks for G02, 57.0 ks for G03, and 60.0 ks for G04, respectively. We examined those images in the 1-5 keV band, where the signal to noise ratio is expected to be optimal owing to low instrumental background. None of the targets were detected (Fig. \ref{fig:e15imgs}). With the point spread function (PSF) of the ACIS-S, 60\% of counts from a point source are expected to fall within a 1-pixel ($=0.49\arcsec$) radius (MARX document 5.2.2). No counts were observed within a 1-pixel radius from any of the four target positions. The background counts in the 1-5 keV band are 0.013-0.015 cts pixel$^{-1}$ for these observations. Therefore no background count is expected (the probability of observing one or more background counts in the area is 4\%), as observed. Detection of no counts sets the 95\% upper limit to be 3 counts in each aperture, using a Bayesian formulation for a Poisson distribution of counts \citep{kraft1991}. Assuming that each source has a photon-index $\Gamma=2$ \citep[e.g.][]{huang2020} and absorbing column \nH\,$<10^{23}$ \psqcm, the corresponding rest-frame 2-10 keV luminosity at the median redshift of the four targets, $z=6.13$, is $2\times 10^{44}$ \ergps, which we adopt as the upper limit for each object for the discussion below.

We note that one count was registered at 0.8 arcsec (within a 2-pixel radius) from the target position in the G03 image. Given the PSF, the probability of detecting one source count in the second radial pixel but none in the first is 10\%. Given the background level, observing one background count within the same area has a probability of 12\%, assuming a Poisson process. Since the Bayes factor is around unity, no preference is given whether it is a source or background count. Even if it is a source count, the one count does not constitutes a detection, but gives an increased upper limit of 4.7 counts for G03, corresponding to $3\times 10^{44}$ \ergps.

\begin{figure}
\centerline{\includegraphics[width=0.4\textwidth,angle=0]{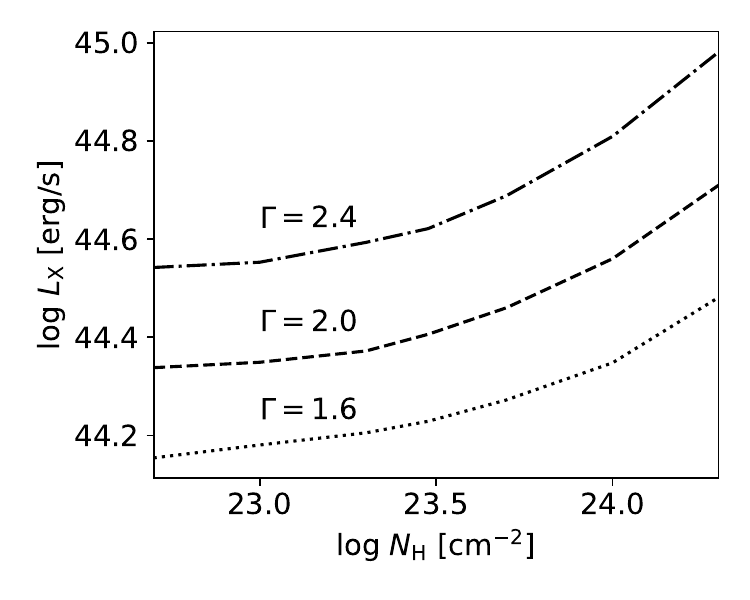}}
\caption{Upper limit of intrinsic $L_{\rm X}$ corresponding to 3 counts (the 95\% upper limit obtained from a 60 ks exposure) for different assumed spectral slope $\Gamma $. $L_{\rm X}$ is the unabsorbed, rest-frame 2-10 keV luminosity. Examples for three values of photon index $\Gamma =1.6$ (dotted line), 2.0 (dashed line) or 2.4 (dot-dashed line) as a function of \nH\ are shown.}
\label{fig:uplim}
\end{figure}

If the spectral slope differs from $\Gamma = 2$ and the absorbing column \nH\ is larger than $10^{23}$ \psqcm, the upper limit of intrinsic (unabsorbed) $L_{\rm X}$ will be different. Fig. \ref{fig:uplim} shows the 95\% upper limits on $L_{\rm X}$ corresponding to 3 counts, when photon index is $\Gamma = 1.6$, 2.0 or 2.4, as a function of absorbing column density, \nH.

We have four targets with approximately equal exposure times. If our four targets have similar X-ray luminosities, as in the case for Ly$\alpha $ --- the mean is $10^{44.1}$ \ergps\ and the standard deviation in the log is only 0.1 ---,  their average X-ray luminosity can be constrained by stacking all the observations. When a 2-pixel radius around the target position is adopted and the one count registered in all the observations is considered, a 95\% upper limit of 1.0 count per object is obtained following the formulations of \citet{weisskopf2007}. For $\Gamma = 2$, this corresponds to $L_{\rm X} = 0.8\times 10^{44}$ \ergps, which we consider as the upper limit of the average $L_{\rm X}$ from stacking.

\section{Discussion}

The fact that we did not detect X-rays from these sources suggests the possibility that these objects are not AGN, and star formation is responsible for the luminous Ly$\alpha $. If that is the case, they would have low metallicity of $Z<0.01Z_{\sun}$ and a young age (less than $10^6$ yr) of intense star formation \citep{hashimoto2017}, since normal star forming galaxies cannot produce EW(Ly$\alpha $)$>240 $\AA \ \citep{malhotra2002}. 
However, recently acquired JWST spectra of the Chandra targets favour presence of AGN for all four objects \citep{matsuoka2025}. We briefly discuss the implications of the JWST results and, if AGN exist, how their X-ray emission remains below the detection limit.

\subsection{Selection bias}

One of the selection criteria of quasar candidates from the Subaru HSC imaging data is that a candidate source to be point-like\footnote{Extended sources are also selected as high-redshift quasar cadidates, but they are given lower priority for spectroscopic follow-up.} \citep{matsuoka2018pasj}, which optimises the original purpose of selecting unobscured quasars in which AGN emission dominates the rest-frame UV light. On the other hand, the HSC image of an obscured AGN is primarily due to emission from the host galaxy, and thus could be extended. This would mean that some obscured AGN candidates would be  missed, and the true number of obscured quasars could be larger than actually discovered under the adopted criteria. In Sect. 2.2, we made the crude estimate of the typical $L_{\rm X}$ of our Chandra targets using the targets' surface density and an assumed luminosity function as a function of the ratio of obscured to unobscured AGN ($\beta = 1$-10). The possible underestimate of obscured quasars mentioned above would then lower the typical luminosity of the obscured quasars that can be selected in the Subaru HSC survey. This could be consistent with no detection of the Chandra targets, but a question is the extent to which we missed obscured quasars.

The upper limit value of the average luminosity for our Chandra targets obtained above (corresponding to $M_{1450}\sim -24$) gives a number of obscured quasars five times larger for $\beta=1$ or $\sim 100$ times larger for $\beta=10 $ than those selected for our Chandra observations (that is, five, including the one left out), using the shape of the luminosity function of \citet{matsuoka2018lf}. Given the negative slope of the luminosity function, these are lower limits of the obscured quasar number. While missing a factor of four more obscured quasars by applying the point source criterion, as found for $\beta=1$, might just be possible, missing by $\sim 100$ times more objects ($\beta = 10$) seems unlikely. On the other hand, a large value of $\beta$, close to $\beta = 10$, is generally favoured, as argued in Sect. 1. This means that the selection bias against obscured quasars is insufficient by a large margin to explain the null X-ray detection.

\subsection{Implications from JWST spectroscopy}

\begin{table*}
  \caption{Results from JWST-NIRSpec observations. }
  \label{tab:jwst}
  \centering
  \begin{tabular}{ccccccccc}
    \hline\hline
    Name & $A_{\rm V}^{\rm N}$ & $A_{\rm V}^{\rm B}$ & $L$(H$\alpha ^{\rm N})$ & $L$(H$\alpha ^{\rm B})$ & $L$([O{\sc iii}]) & $L_{5100}^{\rm obs}$ & $L_{\rm bol}$ & EW(H$\alpha ^{\rm B}$)\\
    \hline
    G01 & $0.10 \pm 0.04$ & $2.15\pm 0.56$ & $43.29\pm 0.01$ & $43.43\pm 0.18$ & $43.64\pm 0.02$ & $44.84\pm 0.25$ & $46.42\pm 0.02$ & $580\pm 160$ \\
    G02 & $0.46 \pm 0.06$ & $1.01\pm 0.06$ & $43.42\pm 0.02$ & $44.04\pm 0.02$ & $43.86\pm 0.03$ & $44.94\pm 0.03$ & $46.64\pm 0.03$ & $878\pm 30$ \\
    G03 & $0.94 \pm 0.05$ & $1.40 \pm 0.17$  & $43.72\pm 0.02$ & $43.70\pm 0.06$ & $43.99\pm 0.02$ & $44.91\pm 0.08$ & $46.77\pm 0.02$ & $557\pm 27$ \\
    G04 & $1.75 \pm 0.08$  & $1.88 \pm 0.07$ & $43.63\pm 0.03$ & $44.36\pm 0.02$ & $44.00\pm 0.04$ & $45.18\pm 0.03$ & $46.78\pm 0.04$ & $1356\pm 26$ \\
    \hline
    \end{tabular}
  \tablefoot{Extract from JWST-NIRSpec spectroscopy results from \citet{matsuoka2025}. Extinction, $A_{\rm V}$, of NLRs (N) and BLRs (B), derived from Balmer decrement. Luminosity of narrow and broad H$\alpha $ emission, [O{\sc iii}]$\lambda 5007$ emission, and continuum at 5100\,\AA. $L$(H$\alpha ^{\rm N}$) and $L$([O{\sc iii}]) have been corrected for the NLR extinction, while $L$(H$\alpha ^{\rm B}$) and $L_{5100}^{\rm obs}$ corrected for the BLR extinction. $L_{\rm bol}$ is bolometric luminosity estimated from $L$([O{\sc iii}]). All luminosities are in logarithmic values and in units of \ergps. Rest-frame EW of broad H$\alpha $ in units of $\AA $.}
\end{table*}

\subsubsection{Broad Balmer emission}

Rest-frame optical spectra of the four Chandra targets have been obtained using JWST-NIRSpec during Cycle-2 (Program ID 3417). All four spectra show weak broad Balmer emission with FWHM$\sim $2000 km\,s$^{-1}$ at the base of narrow lines, reminiscent of those of JWST-selected AGN. Detailed results are presented in \citet{matsuoka2025}. Here we present an extract of emission-line properties relevant to this work in Table \ref{tab:jwst}. Extinction was derived from Balmer decrements of narrow and broad components separately, and luminosities have been corrected for the corresponding extinctions. As broad emission is seen both in H$\alpha $ and H$\beta $, they resemble optical spectra of Seyfert 1.8 \citep[][although J1254--0014 with stronger broad emission than the others barely qualifies as Seyfert 1.5 when the criterion of \citet{winkler1992} is applied]{osterbrock1977}. A straightforward interpretation of the broad Balmer emission is that it originates from the BLRs of AGN.

\subsubsection{Obscured BLRs and continuum source}

The Balmer line profiles with a strong narrow core shows that the broad emission in these objects is systematically weaker than those of typical broad-line quasars. In addition to small to moderate extinction for the NLRs, the visibility of broad H$\beta $ means that extinction for the BLRs cannot be large and is insufficient to suppress the broad H$\alpha $ as observed (Table \ref{tab:jwst}).
This implies either that the BLR emission is intrinsically weak, or only part of the BLRs is visible. The former could occur for a low Eddington ratio $\lambda_{\rm Edd} < 10^{-2}$ \citep[e.g.][]{elitzer2009,kang2024}. However, in this case, it would be accompanied by a hard SED, that is, X-ray emission that is strong relative to UV \citep{ferland2020}.
Since this does not match the observations, this hypothesis seems unlikely. Moreover, given the large bolometric luminosities (see Fig. \ref{fig:maiolino}) their inferred black hole masses would be close to or in excess of $10^{10}\,M_\sun$, which is unlikely at $z\sim 6$ \citep[e.g.][]{wu2015}. More probable is the latter model, in which the bulk of the BLRs is totally hidden. If an optically-thick obscuring torus
is present, a part of the BLRs can be hidden at intermediate viewing angles, and strength of broad-line emission depends on the visiblility of the BLR clouds.

We next make a crude estimate of the total broad H$\alpha $ luminosity and its visible proportion. For broad-line AGN at low redshift, \citet{jin2012a} found that the narrow component of H$\alpha $ emission\footnote{\citet{jin2012a} modelled Balmer lines with three components. As their narrow component matches the profile of [O{\sc iii}], it corresponds to our narrow component, as defined in \citet{matsuoka2025}.} comprises, on average, $10\pm 8$\% of the total flux. Assuming the same composition, a broad component luminosity, $L^{\prime}$(H$\alpha ^{\rm B})$, that includes the hidden portion, can be estimated, using the observed narrow H$\alpha $ luminosity (Table \ref {tab:hidden}). Comparing this estimate with the observed $L$(H$\alpha ^{\rm B}$) gives the fraction of the BLRs which is visible, $f_{\rm BLR}$, ranging from 9\% to 52\% (Table \ref{tab:hidden}).

\begin{figure}
\centerline{\includegraphics[width=0.4\textwidth,angle=0]{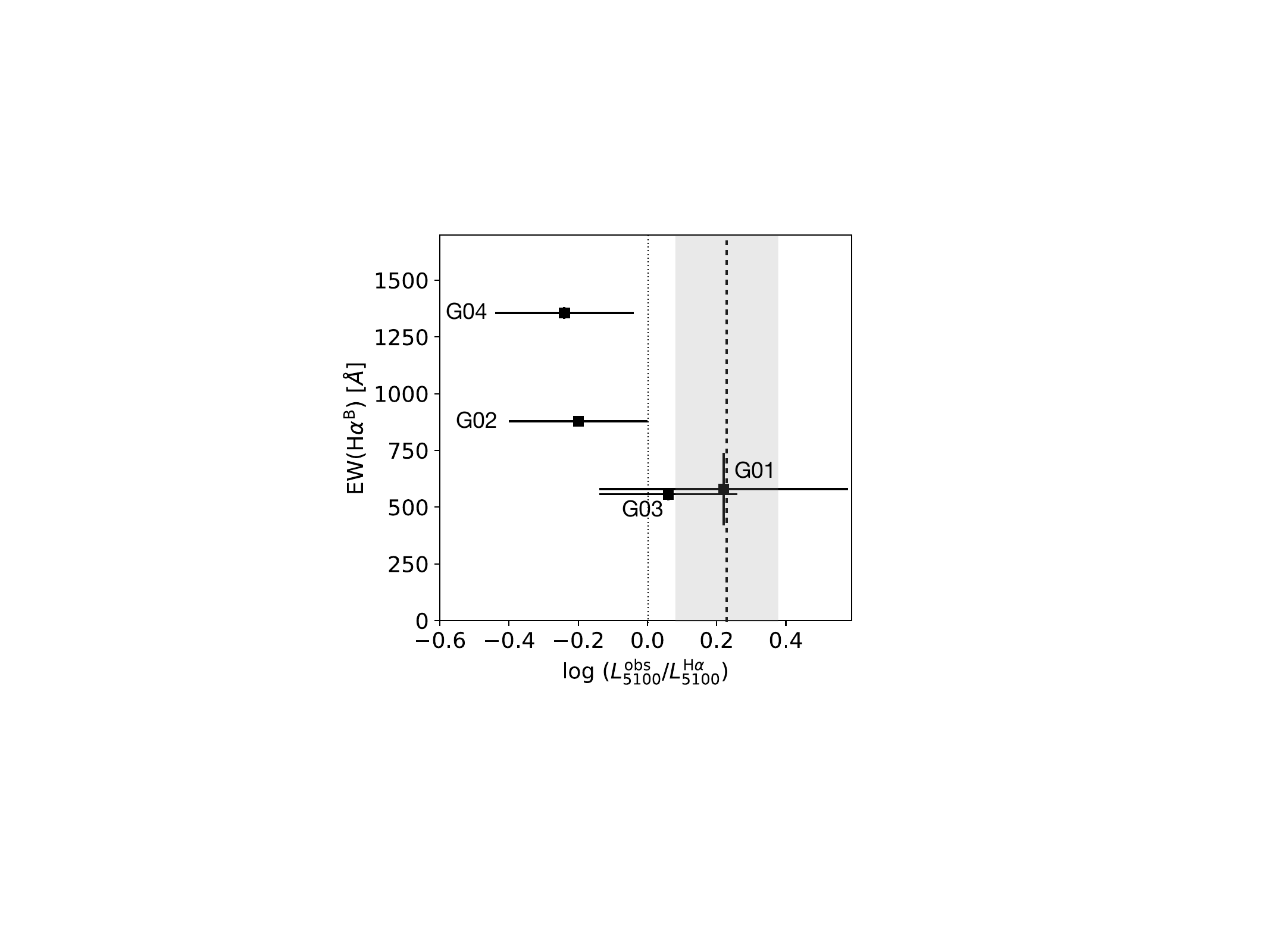}}
\caption{Rest-frame EW of broad H$\alpha $ in units of \AA\ as a function of optical continuum deficit at $5100\,\AA$ given the broad H$\alpha $ luminosity, represented by log ($L_{5100}^{\rm obs} / L_{5100}^{{\rm H}\alpha }$). The dotted line indicates the empirical $L_{5100}$-$L$(H$\alpha ^{\rm B}$) relation by \citet{greene2005} for SDSS AGN at $z<0.35$. The dashed line and the shaded area show log ($L_{5100}^{\rm obs} / L_{5100}^{{\rm H}\alpha })=0.22$, updated by data for luminous quasars at $z=6$-7 and the 68\% interval of posterior predictives, respectively.} 
\label{fig:ewHaB}
\end{figure}

The JWST spectra also show faint optical continuum emission. 
Given a broad H$\alpha $ luminosity, the empirical relation of \citet{greene2005} gives a 5100\,\AA\ luminosity, $L_{5100}^{{\rm H}\alpha}$, with uncertainty $\simeq 0.2$ dex.
The observed luminosity, $L_{5100}^{\rm obs}$ (corrected for $A_{\rm V}^{\rm B}$), is expected to match that obtained from a broad H$\alpha $ luminosity\footnote{We note that the luminosity of broad H$\alpha $ used here is the observed one (Table \ref{tab:jwst}), not $L^{\prime}$(H$\alpha $) discussed above.}, $L_{5100}^{{\rm H}\alpha}$, or larger as there is a contribution from host galaxy light in the observed continuum. While two faint broad Balmer-line objects, G01 and G03, show comparable $L_{5100}^{\rm obs}$ to the expected values, the G02 and G04 with stronger broad Balmer emission show $L_{5100}^{\rm obs}$ fainter than expected (Fig. \ref{fig:ewHaB}), suggesting that the AGN continuum is suppressed more than the BLR emission is. Although the \citet{greene2005} relation is based on SDSS broad-line AGN at lower redshift ($z<0.35$), we verified that the relative continuum deficit is not an inherent feature of high-redshift quasars albeit for a small sample. We looked into $L_{5100}^{\rm obs}/L_{5100}$ of luminous, unobscured quasars at $z=6$-7 observed with JWST \citep{marshall2023,loiacono2024,bosman2024,yue2024_eiger,marshall2025}. These high-redshift quasars instead all show positive values of log ($L_{5100}^{\rm obs}/L_{5100}^{{\rm H}\alpha})\approx 0.1$-0.4, while any host galaxy contribution in these quasars should be negligible. These positive offsets could be understood by the Baldwin effect \citep{baldwin1977}, as they are more luminous than the objects in the \citet{greene2005} sample, rather than a redshift evolution. The updated reference log ($L_{5100}^{\rm obs}/L_{5100}^{{\rm H}\alpha}$) value for unobscured high-redshift quasars, using these observations, is indicated in Fig. \ref{fig:ewHaB} with a predicted 68\% scatter interval. SHELLQs broad-line quasars observed with JWST also show similar positive offsets (M. Onoue, priv. comm.).  Therefore the continuum deficit relative to the BLR emission in the Chandra sources appears to be supported.
The observed large EWs of broad H$\alpha $ (Table \ref{tab:jwst}) are in accord with this conclusion. These EWs from 557\,\AA\ to 1356\,\AA\ are larger than in low-redshift SDSS quasars that typically have EW\,$\sim 200\,\AA$, and similar to values for JWST-selected AGN \citep{maiolino2025}. Negative values of log ($L_{5100}^{\rm obs}/L_{5100}^{{\rm H}\alpha}$) and large EWs of broad H$\alpha $, as discussed above, also hold for the other five SHELLQs narrow-line objects with broad H$\alpha $ detection of \citet{matsuoka2025}.
If part of the BLRs is hidden, as we argued, the continuum source in our objects should be totally hidden, too, due to its inner location, and the observed continuum could be dominated by host galaxy light, apart from possible scattered light or leaked light if obscuration is patchy (related discussion can be found in \citet{matsuoka2025}).

\begin{table}
  \caption{Correction for hidden BLRs and $L_{\rm X}$ predictions}
  \label{tab:hidden}
  \centering
  \begin{tabular}{ccccc}
    \hline\hline
    Name & $L^{\prime}$(H$\alpha ^{\rm B})$ & $f_{\rm BLR}$ & $L^{\prime {\rm H}\alpha^{\rm B}}_{\rm X}$ & $L^{\rm OIII}_{\rm X}$ \\
     & (1) & (2) & (3) & (4) \\
    \hline
    G01 & 44.24 & 0.15 & 45.07 (45.00) & 44.96 \\
    G02 & 44.30 & 0.54 & 45.12 (45.03) & 45.16 \\
    G03 & 44.53 & 0.19 & 45.31 (45.16) & 45.22 \\
    G04 & 44.58 & 0.59 & 45.35 (45.18) & 45.32 \\
    \hline
    \end{tabular}
  \tablefoot{(1) Broad H$\alpha $ luminosity, corrected for the fraction of the BLR that is hidden (see text in Sect. 5.2.3). (2) Visible fraction of the BLRs. (3) 2-10 keV luminosity inferred from the $L_{\rm X}$-$L$(H$\alpha $) relation of \citet{jin2012a} and $L^{\prime}$(H$\alpha^{\rm B}$) in column (1). Those in parentheses are values corrected for a possible over-estimate that originates from an extrapolation of the relation (see text in Sect. 5.2.3). (4) 2-10 keV luminosity, estimated from the typical absorption-corrected X-ray to extinction-corrected [O{\sc iii}]$\lambda 5007$ ratio of log $(L_{\rm X}/L([$O{\sc iii}$])=1.32$. Luminosities are logarithmic values in units of \ergps. Uncertainty of these luminosities comes from the scatter of the relations adopted, about 0.3 dex.}
\end{table}

\subsubsection{X-ray quietness}

\begin{figure}
\centerline{\includegraphics[width=0.4\textwidth,angle=0]{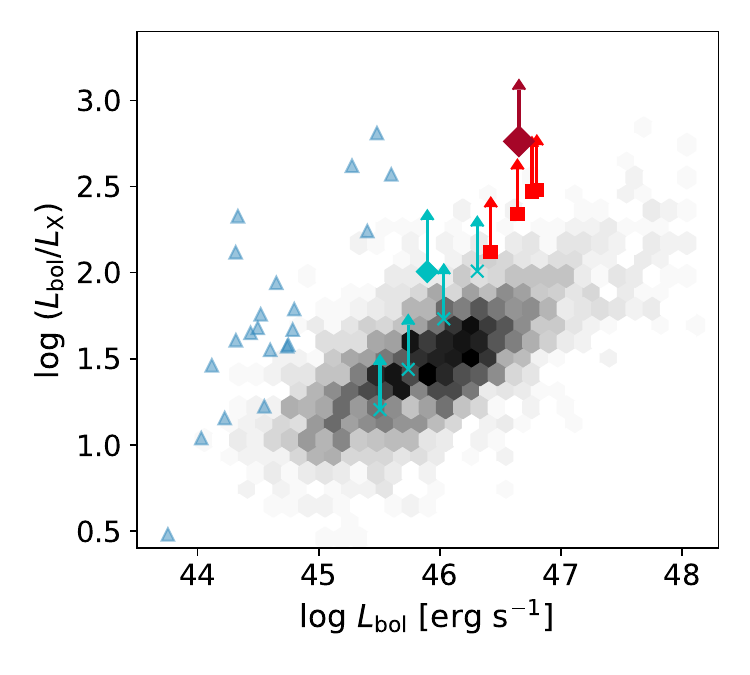}}
\caption{X-ray to bolometric luminosity ratio ($L_{\rm bol}/L_{\rm X}$) against $L_{\rm bol}$. The four Chandra targets are indicated by red squares. The diamond symbol shows the result of stacking the four. Cyan crosses are those for the individual objects (diamond for the stack) where $L_{\rm bol}$ was estimated from broad H$\alpha $ (see Table \ref{tab:jwst}). Blue triangles indicate JWST-selected AGN from \citet{maiolino2025}, where $L_{\rm bol}/L_{\rm X}$ values are all lower limits, as X-ray emission was detected from none of these sources. Data for broad-line quasars \citep{lusso2020} are shown in grey-scale two-dimensional histogram.}
\label{fig:maiolino}
\end{figure}

Assuming the hypothesis that the BLRs in these sources are partially hidden, the observed broad H$\alpha $ luminosity depends on the fraction of the BLRs directly visible and the true luminosity should be larger than the observed value. Thus we re-evaluate the broad H$\alpha $-based $L_{\rm X}$ of AGN, given our knowledge of low-redshift broad-line AGN. With the corrected H$\alpha $ luminosities, $L^{\prime }$(H$\alpha $), the $L_{\rm X}$-$L$(H$\alpha $) relation \citep{jin2012a} predicts X-ray luminosities ($L^{\prime}_{\rm X}$) exceeding $10^{45}$ \ergps\ for all the Chandra targets (Table \ref{tab:hidden}). We note that, as those $L^{\prime }$(H$\alpha $) are above the luminosity range ($\leq 10^{44}$ \ergps\ in H$\alpha $) where the $L_{\rm X}$-$L$(H$\alpha $) relation was derived, an extrapolation of the relation may overestimate $L_{\rm X}$, since $L_{\rm X}/L_{\rm UV}$ decreases towards higher luminosities. With the known $L_{\rm X}$-$L_{\rm UV}$ relation for quasars \citep[e.g.][see also Fig. \ref{fig:maiolino}]{vito2019,lusso2016}, we found the possible over-estimate in $L_{\rm X}^{\prime }$ to be 0.07 dex to 0.17 dex. Even taking this into account, the predicted $L_{\rm X}$ are still found to be $10^{45}$ \ergps\ or larger (Table \ref{tab:hidden}). 

Similar $L_{\rm X}$ values are also expected from the observed [O{\sc iii}]$\lambda 5007$ luminosity, which is considered to be a reliable indicator of isotropic AGN power \citep[e.g.][]{kauffmann2003,shen2011,rakshit2020}, especially when the BLRs are obscured as in Type 2 AGN.
With the large observed luminosities of our Chandra targets, any contribution of star formation to the narrow emission-lines should be minor otherwise the continuum would be much brighter than observed (see Sect. 5.2.2).
In particular, [O{\sc iii}] is correlated with absorption-corrected $L_{\rm X}$. The luminosity ratio of 2-10 keV X-ray to [O{\sc iii}] measured for low-redshift AGN is log ($L_{\rm X}/L$([O{\sc iii}])$\approx 1.9$ \citep{mulchaey1994,heckman2005,trouille2010,georgantopoulos2010,jin2012a,stern2012b,berney2015,malkan2017}. The scatter of this ratio from various paper ranges from 0.3 dex to 0.5 dex. These values are derived, however, with [O{\sc iii}] luminosities with no extinction correction. \citet{jin2012a} obtained one of the typical values, 1.88, which reduces to 1.32 when extinction correction from the mean Balmer decrement (H$\alpha $/H$\beta = 5.0$) for their broad-line AGN sample is applied. Using this value, our Chandra targets are expected to have $L_{\rm X} = ($1-2$) \times 10^{45}$ \ergps (Table \ref{tab:hidden}).

The agreement between the predicted $L_{\rm X}$ (and hence bolometric luminosities, $L_{\rm bol}$) by two independent estimators, $L$([O{\sc iii}]) and corrected $L^{\prime}$(H$\alpha^{\rm B}$), gives some credence to the assumption of partially hidden BLRs.
Fig. \ref{fig:maiolino} shows the bolometric to X-ray luminosity ratio, $L_{\rm bol}$/$L_{\rm X}$ (or X-ray bolometric correction), against $L_{\rm bol}$ for the Chandra targets on the diagram, similar to Fig. 2 of \citet{maiolino2025}. Two sets of data points for the Chandra targets using $L_{\rm bol}$, estimated either from $L$([O{\sc iii}]) \citep[red squares,][Table 7]{matsuoka2025} or from broad H$\alpha$ (cyan crosses)\footnote{First, AGN continuum luminosity at 5100\,\AA, $L_{5100}^{{\rm H}\alpha}$, was estimated, using the empirical relation \citep{greene2005} with extinction-corrected broad H$\alpha $ luminosity, and then converted to $L_{\rm bol}$ with the bolometric correction of 7.79 \citep{krawczyk2013}.} luminosities, are shown. For comparison, data for 2421 broad-line quasars at $z=0.01$-7 (mostly $z\leq 3$) studied by \citet{lusso2020} are shown in the grey-scale two-dimensional histogram\footnote{$L_{\rm bol}$ was derived from monochromatic luminosity at 2500\,\AA\ given in \citet{lusso2020} with the bolometric correction of 2.75 \citep{krawczyk2013} while $L_{\rm X}$ was derived from the 2 keV luminosity assuming $\Gamma = 2$ as the spectral slope in the 2-10 keV band.} as well as the data for JWST-selected AGN at $z=$\, 4-7 \citep{maiolino2025}, all of which are lower limits, as indicated by blue triangles\footnote{$L_{\rm bol}$ of the JWST-selected AGN were based on broad H$\alpha $ but have been adjusted by using the same bolometric correction as used for the Chandra sources. $L_{\rm X}$ values have also been adjusted by matching the same assumed spectral slope of $\Gamma = 2$, as used for our Chandra sources and the broad-line quasars ($\Gamma = 1.7$ was assumed in \citet{maiolino2025}).}.
We make a cautionary note that, if the Chandra targets indeed share similar characteristics with LRDs, their bolometric luminosity estimates based on the conventional means could be highly uncertain, as discussed by e.g. \citet{greene2025}.

The luminosities of strong narrow lines are similar between the four objects: the logarithmic means and standard deviations of the luminosities of narrow H$\alpha $ and [O{\sc iii}] in units of \ergps\ are 43.59 (0.19) and 43.83 (0.10). The small dispersion of these line luminosities leads to similar predicted $L_{\rm X}$ (Table \ref{tab:hidden}) and $L_{\rm bol}$ (Fig. \ref{fig:maiolino}, Table \ref{tab:jwst}). Incidentally, these values agree with the estimates in Sect. 2.2 based on the surface density with a large $\beta\sim 10$. 

The X-ray upper limits obtained from Chandra observations are nearly one order of magnitude (1.5 dex for the stack) below the expected $L_{\rm X}$ argued above (see also Fig. \ref{fig:maiolino}). We discuss possible causes of the X-ray quietness of these objects below.

\subsection{Possible explanations of X-ray quietness}

The X-ray quietness of these objects can be due either to intrinsically weak X-ray emission or to heavy obscuration. The former possibility is linked to a steep UV to X-ray SED in a super-Eddington accretion flow, as discussed for the JWST-selected AGN \citep{inayoshi2024b,maiolino2025}. Here we discuss the latter. The Chandra bandpass we observed (1-5 keV) corresponds to 7-35 keV in the rest frame. Suppressing this hard X-ray emission requires a significantly Compton-thick absorbing column of \nH $>> 10^{24}$\,\psqcm.

\subsubsection{Compton-thick absorption by interstellar medium}

In the hypothesis of partially hidden BLRs argued above, the X-ray source is located behind the same obscuring matter at intermediate viewing angles. Although this picture is similar to the unification scheme of Seyfert galaxies with a pc-scale torus \citep{antonucci1985}, nuclear obscuration may take a different form at high redshifts, as argued by \citet{gilli2022}. With the increased gas content in galaxies \citep[e.g.][]{scoville2017}, a galaxy-wide interstellar medium (ISM) is capable of making up a Compton-thick absorbing column over the bulk of solid angles. In fact, ALMA high-resolution maps of high-$J$ transition CO lines of a quasar at $z=6$ showed that warm molecular gas extends to 100-pc scales, for which a column density of \nH $\simeq 10^{25}$ \psqcm\ is inferred \citep{tadaki2025}. Given the small host galaxy size of the order of 1 kpc \citep[][Ding et al. in prep.]{ding2023,onoue2024}, this dense gas occupies a significant portion of the host galaxy. Even though the geometry is not yet clear, this form of Compton-thick gas can completely suppress X-ray emission in the Chandra targets, as well as the optical emission from inner radii.

Such extended obscuring matter has a thermal structure that is likely to deviate from that in the typical hot dust seen in low-redshift AGN \citep{barvainis1987}. This may, in part, be similar to an extended dusty region, proposed to explain the lack of a near-infrared bump in the rest frame in LRDs \citep{li2025}. Photometric observation of the rest-frame near-infrared band for these SHELLQs narrow-line objects can probe the nature of the obscuring matter.

\subsubsection{Obscuration with a supercritical accretion flow}

If the idea of partially hidden BLRs is discounted and only mild obscuration, as implied from the values of extinction measured from the Balmer decrement (Table \ref{tab:jwst}), is present for the optical emission region in these objects, then a distinct Compton-thick obscuration that solely affects the X-ray source is required. Various explanations have been proposed for the X-ray quietness of JWST-selected AGN, involving optically-thick gas within the dust-sublimation radius \citep[e.g.][]{maiolino2025,deugenio2025,ji2025,naidu2025,inayoshi2025}. We consider here a supercritical accretion flow as one of possible mechanisms of strong X-ray attenuation using a model developed for ultra-luminous X-ray sources \citep[ULXs,][see also \citet{king2025}]{kawashima2009,kitaki2017,kawanaka2021,ogawa2021} when combined with an orientation effect, as discussed below.

In the disc accretion mode, as opposed to spherical accretion, matter falls along the disc plane while radiation can escape in the polar direction, enabling a supercritical accretion flow while emitting at super-Eddington luminosity \citep[e.g.][]{ohsuga2005}. A number of key features of such an accretion flow are described in the slim disc prescription by \citet{abramowicz1988}. One of the relevant features in our argument is an optically thick funnel created by the innermost part of the disc, which is inflated by strong radiation pressure. It collimates the disc radiation, making it anisotropic \citep{ohsuga2005,ohsuga2007,takahashi2015,sadowski2015,jiang2019}. In AGN, the disc radiation is in the UV range and an external mechanism is needed to produce X-rays. \citet{kawashima2009,kawashima2012} proposed that the radiation-pressure driven disc winds in a supercritical accretion flow can act as a Comptonising corona that produces X-rays \citep[see also][]{jiang2019}. The dynamical nature makes this wind-fed corona somewhat cooler and more optically thick than the conventional corona \citep[e.g.][]{haardt1991} above the standard thin disc in low-redshift AGN \citep{kawanaka2021,inayoshi2024b}. In this case, the X-ray emission is also anisotropic. We refer to \citet{kawashima2012, kitaki2017,ogawa2021} for details of the viewing-angle dependence of X-ray luminosity and spectral shape.

Viewing such a system face-on, the whole X-ray emitting region, including the high-temperature (a few $10^8$ K) part, deep down the funnel, is visible. Ultra-luminous X-ray sources (ULXs) are considered to fit in this situation \citep{kawashima2009,kitaki2017,king2024}, and their AGN couterparts perhaps correspond to the most luminous, unobscured quasars. In quasars, however, disc emission moves to the UV range and the Compton-scattered emission dominates the X-ray range, unlike a stellar-mass black hole that emits X-ray disc emission. Given the cooler wind-fed corona, the X-ray spectra steepen above $\sim 10$ keV, below which they can be as hard as low-redshift AGN that show $\Gamma = 1.9$-2.0, on average \citep[e.g.][]{vignali2003,steffen2006,just2007,lusso2016,nanni2017,huang2020}. Since XMM-Newton or Chandra bandpass mainly covers rest-frame energies higher than 10 keV in $z\geq 6$ quasars, a single power-law fit would yield a steeper slope, as observed in the most luminous quasars at $z\sim 6$ \citep[$\Gamma\simeq 2.2$ and $\Gamma\sim 2.4$ respectively in][] {vito2019,zappacosta2023}. The fact that the spectral steepening is only found for high-redshift quasars could therefore be a combination of the rest-frame bandpass effect and a selection effect that spectral slope measurements are possible only for the most luminous quasars given the limited sensitivity. Due to the collimation, the X-ray luminosities of those most luminous quasars are likely boosted.

On the other hand, when the system is viewed at some inclined angle, the flux boosting effect disappears and the innermost part of the funnel, where hard X-rays are produced, is blocked from our direct view by the inflated inner disc which is optically thick. Any X-ray emission that is still visible is therefore much weaker and has a softer spectrum than that seen in the face-on case \citep{kitaki2017,ogawa2021}. Those luminous quasars at $z\sim 6$, such as those in the HYPERION sample \citep[][see also \citet{vito2019}]{zappacosta2023} emit at $L_{\rm X}\sim 2\times 10^{45}$ \ergps\ on average. If they are the face-on version of AGN with a supercritical accretion argued above, the same nuclei viewed at an intermediate angle would have an observed luminosity of a few times $10^{43}$\ergps (for example, the rest-frame 20 keV emission when viewed at $40^{\circ}$ would be $\sim $2 orders of magnitude fainter than when viewed face-on). As this expected $L_{\rm X}$ fits the Chandra upper limit, the SHELLQs narrow-line objects could contain nuclei similar to the HYPERION quasars but viewed at intermediate angles. Thus the inherent anisotropy of X-ray radiation at the innermost radii of a supercritical accretion flow offers an alternative explanation of the X-ray quietness of the Chandra targets.

\section{Summary}

We selected obscured AGN candidates at $z\sim 6$ from the SHELLQs sample for X-ray observations with Chandra, on account of thier luminous Ly$\alpha $ and faint UV continuum. We did not detect X-rays; this, together with subsequent detection of weak broad Balmer emission in the rest-frame optical spectra taken with the JWST-NIRSpec, suggests that they may be luminous counterparts to JWST-selected AGN, including LRDs, discovered at high redshifts. We interepret their weak broad Balmer emission and faint optical continua as a result of the BLRs being partly visible, while the rest of the BLRs, the optical continuum and X-ray emitting regions of the accretion disc are hidden behind an optically-thick obscuring torus. Any X-ray emission can be suppressed if the obscuring matter is Compton-thick, due to the ISM of the host galaxies, as hypothesised and observed at high redshifts. Alternatively, an inflated inner accretion disc in a supercritical accretion flow could strongly suppress X-ray emission with an orientation effect. 

\begin{acknowledgements}
We thank Shin Mineshige, Livia Vallini, Roberto Decarli for useful discussion, and Guido Risaliti for providing us with the data used for Fig. \ref{fig:maiolino}. This paper employs a list of Chandra datasets, obtained by the Chandra X-ray Observatory, contained in the Chandra Data Collection (CDC) 396~ \href{https://doi.org/10.25574/cdc.396}{doi:10.25574/cdc.396} and made use of software provided by the Chandra X-ray Center (CXC). This work is based in part on observations made with the NASA/ESA/CSA James Webb Space Telescope. The data (DOI: 10.17909/cve4-9p26) were obtained from the Mikulski Archive for Space Telescopes at the Space Telescope Science Institute, which is operated by the Association of Universities for Research in Astronomy, Inc., under NASA contract NAS 5-03127 for JWST. These observations are associated with program 3417. Support for this work was provided by NASA through Chandra Award Number 23700304 issued by CXC. KIw acknowledges support under the grant PID2022-136828NB-C44 provided by MCIN/AEI/10.13039/501100011033/FEDER, UE. MO acknowledges support from Japan Society for the Promotion of Science (JSPS) KAKENHI Grant Number 24K22894. YM was supported by the Japan Society for the Promotion of Science (JSPS) KAKENHI Grant No. 21H04494. KK acknowledges the support by JSPS KAKENHI Grant Numbers 22H04939, 23K20035, and 24H00004. RG and FV acknowledge financial support by the INAF Grants for Fundamental Research 2023.
\end{acknowledgements}

\bibliographystyle{aa} \bibliography{shellqs}{}

\begin{appendix}
  \section{Luminosity relation between X-ray and broad H$\alpha$ emission}

\begin{figure}
    \centerline{\includegraphics[width=0.4\textwidth,angle=0]{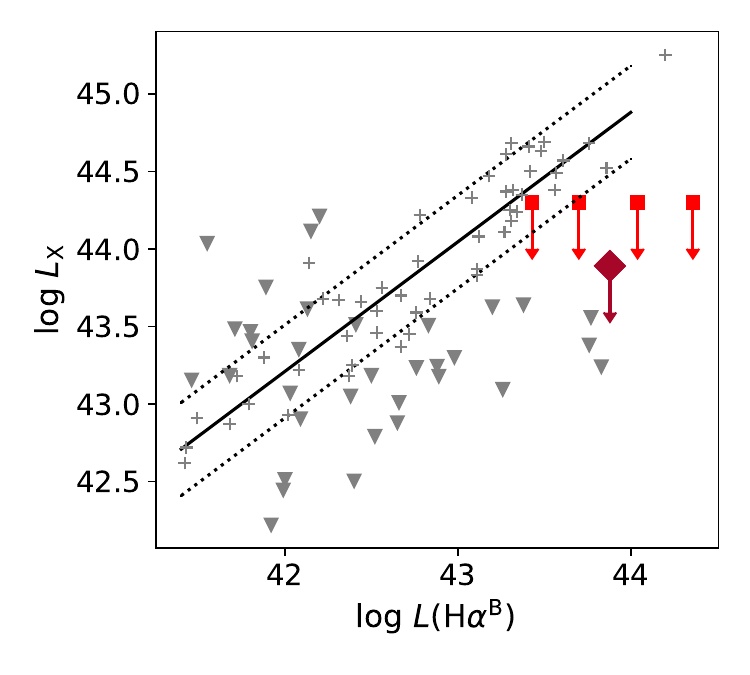}}
      \caption{$L_{\rm X}$ against $L$(H$\alpha $) diagram. The four Chandra targets are indicated by red squares and $L_{\rm X}$ values are 95\% upper limits. The diamond symbol shows the result of stacking the four. Grey triangles indicate LRDs at $z=3$-7 from \citet{yue2024}, when \nH\ $= 10^{22}$ \psqcm\ is assumed. The $L_{\rm X}$ values are also 95\% upper limits, adjusted from the original $3\sigma $ upper limits. The black solid line shows the $L_{\rm X}$-$L$(H$\alpha $) relation for broad-line AGN of \citet[][their formula (1)]{jin2012a} (grey crosses). Dotted lines indicate $\pm 0.3$ dex interval of the relation, corresponding to the data scatter.}
\label{fig:yue}
\end{figure}

In low-redshift unobscured AGN, broad-line H$\alpha $ luminosity correlates with $L_{\rm X}$ and AGN bolometric luminosity, $L_{\rm bol}$ \citep{jin2012a,stern2012a,greene2005,kaspi2000,ward1988}.
Fig. \ref{fig:yue} plots extinction-corrected broad H$\alpha $ luminosity and $L_{\rm X}$ upper limit of the Chandra targets, with the $L_{\rm X}$ - $L$(H$\alpha $) relation derived for broad-line AGN at low redshift\citep{jin2012a} superposed. The H$\alpha $ data of \citet{jin2012a} sample are corrected only for Galactic extinction, but the mean broad H$\alpha $/H$\beta $ ratio (3.1) of the sample suggests little intrinsic BLR extinction. For a comparison, the LRDs at $z=3$-7 \citep{yue2024}, none of which were detected in X-ray, are also plotted.
Even without the geometrical effect on the BLR discussed in the main text, the Chandra upper limits show that our targets might be X-ray weak, similar to LRDs, although the constraints are not as tight as some of the LRDs in the deep X-ray survey fields. We note that the relation could be slightly flatten in the range of $L$(H$\alpha$)$>10^{44}$ \ergps, where some of the Chandra targets lie, because of the decreasing trend of $L_{\rm X}/L_{\rm UV}$ ratio towards higher luminosities (see Sect. 5.2.3 and Fig. \ref{fig:maiolino}).  

  \end{appendix}

\end{document}